\documentclass[10pt]{article}

\usepackage{algorithm}
\usepackage{graphicx}
\usepackage{textcomp}
\usepackage{xcolor}

\usepackage[utf8]{inputenc}
\usepackage{listings}
\usepackage{float}
\usepackage{subfigure}
\usepackage{subcaption}
\usepackage{subfigure}
\usepackage{caption}

\usepackage{amsmath}
\usepackage{hyperref}
\usepackage{cleveref}
\usepackage{tabularx} 

\usepackage{url}
\usepackage{multirow}
\usepackage{cancel}
\usepackage{commath}
\usepackage{algpseudocode}
\usepackage{etoolbox}
\usepackage{varwidth}
\usepackage{tikz}
\usetikzlibrary{shapes,arrows}
\usepackage{colortbl}
\usepackage[margin=1in]{geometry}
\usepackage{booktabs}
\usepackage{tabularx}

\newcolumntype{P}[1]{>{\centering\arraybackslash}m{#1}}

\definecolor{codegreen}{rgb}{0,0.6,0}
\definecolor{codeblack}{rgb}{0.30,0.30,0.30}
\definecolor{codepurple}{rgb}{0.58,0,0.82}
\definecolor{backcolour}{rgb}{0.95,0.95,0.92}
 
\lstdefinestyle{codeListStyle}{
    backgroundcolor=\color{backcolour},
    commentstyle=\color{codegreen},
    keywordstyle=\color{magenta},
    numberstyle=\scriptsize\color{codeblack},
    stringstyle=\color{codepurple},
    breaklines=true,
    breakatwhitespace=true,
    numbers=left,
    xleftmargin=0.2in,
    xrightmargin=0.05in,
    basicstyle=\ttfamily\footnotesize,
    numbersep=5pt,
    frame=single,                  
    tabsize=2
}

\AtBeginDocument{%
  \providecommand\BibTeX{{%
    \normalfont B\kern-0.5em{\scshape i\kern-0.25em b}\kern-0.8em\TeX}}}

\begin{document}

\title{Static Estimation of Reuse Profiles for Arrays in Nested Loops}

\author{Abdur Razzak$^1$,  Atanu Barai$^2$, Nandakishore Santhi$^2$, Abdel-Hameed A. Badawy$^{1,2}$}

\date{
    \small
    ${}^1$ Klipsch School of ECE, New Mexico State University, Las Cruces, NM 80003, USA\\
    ${}^2$ Los Alamos National Laboratory, Los Alamos, NM 87545, USA\\
    \{arazzak, badawy\}@nmsu.edu, \{abarai, nsanthi\}@lanl.gov\\
}

\maketitle

\begin{abstract}
Efficient memory access patterns play a crucial role in determining the overall performance of applications by exploiting temporal and spatial locality, thus maximizing cache locality. The Reuse Distance Histogram (RDH) is a widely used metric to quantify temporal locality, measuring the distance between consecutive accesses to the same memory location. Traditionally, calculating RDH requires program execution and memory trace collection to obtain dynamic memory access behavior. This trace collection is often time-consuming, resource-intensive, and unsuitable for early-stage optimization or large-scale applications. Static prediction, on the other hand, offers a significant speedup in estimating RDH and cache hit rates. However, these approaches lack accuracy, since the predictions come without running the program and knowing the complete memory access pattern. More specifically, when arrays are used inside nested loops, it is quite difficult to predict access patterns of the array references without executing the program. This paper presents a novel static analysis framework for predicting the reuse profiles of array references in programs with nested loop structures, without requiring any runtime information. By analyzing loop bounds, access patterns in smaller problem sizes, and predictive equations, our method predicts access patterns of arrays and estimates reuse distances and cache hit rate at compile time. This paper extends our previous study by incorporating more analysis and improving prediction by addressing previously unhandled reuse patterns. We evaluate our technique against a widely accepted traditional trace-driven profiling tool, Parallel Reuse Distance Analysis (PARDA). The results demonstrate that our static predictor achieves comparable accuracy while offering orders-of-magnitude improvement in the analysis speed. This work offers a practical alternative to dynamic reuse profiling and paves the way for integration into compilers and static performance modeling tools.

\end{abstract}

\textbf{Keywords:} Memory Analysis, Reuse Distance, Static Analysis, Probabilistic Prediction

\section{Introduction}
\label{sec:intro}

Optimizing memory behavior is essential for improving program performance, especially in memory-intensive applications where memory access latency is the primary bottleneck. Among various memory access analysis techniques, RDH has emerged as a powerful and widely used metric to capture the location of temporal data. Obtaining these measurements requires program analysis methods that accurately capture memory access patterns and compute reuse distances (Section~\ref{sec:background}). Traditionally, dynamic profiling techniques are used for analyzing applications with arrays inside loops. These approaches refer to the execution of the program while collecting detailed memory traces to calculate the RDH. This is because without running the application, it is hard to predict the sequence of array references when used inside loops, as it changes the memory address along with the loop variable. Even though this process is accurate, it incurs significant run-time overhead and wait time to generate and process large trace files. The necessity of program execution also makes it difficult to evaluate a program's characteristics across different input sizes or configurations without rerunning the entire profiling process.

In this work, we propose a static analysis tool to predict reuse profiles to circumvent the long trace collection time of dynamic approaches. More specifically, this work predicts the array's volatile reference-changing behavior in nested loops at compile time without the need for trace collection or program execution. 

Figure~\ref{fig:overview} shows the overall workflow for calculating the RDH and the cache hit rate of the source code. We have used the LLVM Static Analysis~\cite{LLVM_Atanu} process for the initial steps until the Loop Annotated Trace. Our Alternative Static Analyzer takes the loop-annotated static traces as input. By analyzing loop bounds, loop nesting structure, and affine array access functions, our approach estimates reuse distances at compile time and generates RDH and cache hit rate of the program. Our static approach, shown in green, bypasses the time-consuming dynamic steps, which are struck through in red, such as running the program and waiting for the run-time trace collection. These are the steps we are avoiding. 
Our technique offers several key benefits:
\begin{itemize}
    \item Enables early-stage performance modeling without running the program.
    \item Facilitates design space exploration independently of the problem sizes.
    \item Significantly reduces the analysis cost compared to trace-driven methods.
\end{itemize}
We validate our static model's reuse distance and cache hit rate predictions by comparing with PARDA~\cite{PARDA:Niu} using a few benchmark applications from Polybench~\cite{polybenchc}. Our model achieves significant accuracy and massive speedup in computation time, underscoring its potential as an efficient alternative to dynamic tools for cache performance modeling.

The remainder of the paper is organized as follows. Section~\ref{sec:background} presents and discusses the concept of reuse distance. Section~\ref{sec:related_works} reviews the current efforts so far in both dynamic and static approaches. Section~\ref{sec:method} explains our static analyzer in detail. Section~\ref{sec:results} introduces the results. Finally, Sections~\ref{sec:limitations} and~\ref{sec:conclusion} discuss the limitations of the study and concludes the paper, respectively.

\section{Background}
\label{sec:background}
Reuse distance (also called LRU stack distance) is a measure of how many unique memory accesses occur between two consecutive accesses to the same memory location. In simpler terms, it tells how far back in time (in terms of unique accesses) a reference was last used before being reused again.

\begin{figure}[ht]
\centering

\begin{tabular}{@{}ccc@{}}
\toprule
\textbf{Index} & \textbf{Access} & \textbf{Reuse Distance (RD)} \\ \midrule
0 & a & $\infty$ (first time) \\
1 & b & $\infty$ (first time) \\
2 & b & 0 (no other ref. between previous b) \\
3 & a & 1 (b) \\
4 & c & $\infty$ (first time) \\
5 & b & 2 (c, a) \\
6 & a & 2 (b, c) \\
\bottomrule
\end{tabular}
\caption*{a. Reuse Distance Table}

\vspace{1em} 

\includegraphics[width=0.55\textwidth]{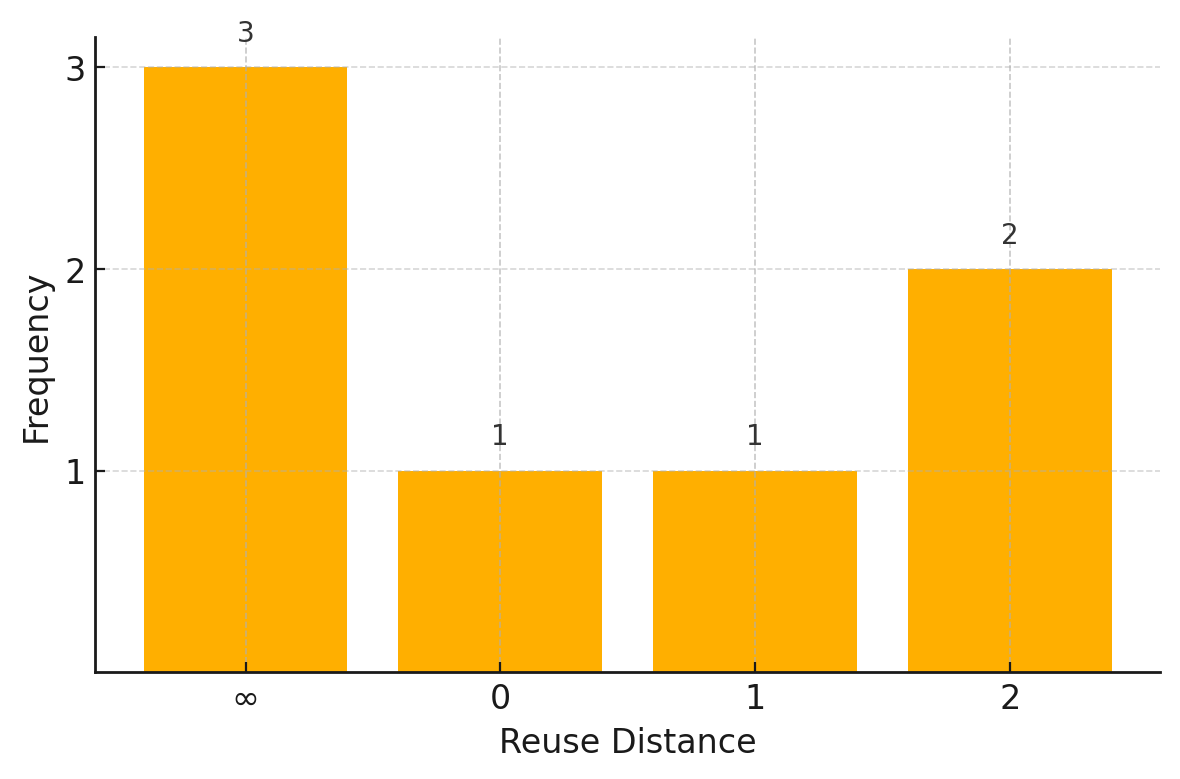}
\caption*{b. Reuse Distance Histogram}

\caption{Reuse distance analysis for example reference accesses: table (a) and histogram (b).}
\label{fig:reuse_distance_analysis}
\end{figure}

Figure~\ref{fig:reuse_distance_analysis} (a) shows a demonstration of the reuse distance calculation from an example memory access sequence: a b b a c b a. When a memory reference enters the stack for the first time, those accesses are certainly cold misses and hence marked as $\infty$. Every other access finds a reuse of the same reference earlier in the memory stack and counts the unique references in between as the reuse distance. RDHs quantify how frequently a memory address is reused, which represents a histogram that shows the occurrences or frequencies of those reuse distance numbers (Figure~\ref{fig:reuse_distance_analysis}(b)). The higher the frequency at lower reuse distances, the better, as these values are more likely to be found in the cache, resulting in cache hits.




\section{Related Work}
\label{sec:related_works}
Reuse distance analysis and cache hit rates have been extensively studied in recent times as a means of understanding and optimizing cache performance. Existing work can be broadly classified into two approaches. Dynamic approaches rely on runtime information and trace collection. Static approaches, on the other hand, aim to predict reuse patterns at compile time without executing the program.

\subsection{Dynamic Reuse Profiling}
Dynamic analysis tools measure reuse distances by tracking memory access traces during program execution. These tools require the program to run first and then full memory accesses after execution are required to begin the calculation process. Some notable such tools are DineroIV~\cite{dineroiv}, Valgrind~\cite{valgrind}, and Pin-based profilers, which instrument the code to capture memory reference behavior and generate accurate RDH. Zhong et al.~\cite{locality_zhong} introduced the concept of temporal reuse distance and developed efficient techniques for runtime profiling. In recent work, including PARDA~\cite{PARDA:Niu}, improves scalability and performance by parallelizing the histogram construction process using a tree-based technique. ReuseTracker~\cite{ReuseTracker} represents a hardware-assisted, low-overhead reuse distance analysis tool for multithreaded programs. They accurately profile both private and shared cache reuses, overcoming the limitations of traditional simulators and instrumentation-based tools by leveraging performance monitoring units (PMUs) and debug registers in commodity CPUs.

Although dynamic profiling remains the gold standard in terms of accuracy, it imposes substantial overhead in both time and calculation storage, making it unsuitable for large-scale or design-time analysis.

\vspace{-1em}

\subsection{Static Reuse Profiling}
In contrast, static analysis techniques estimate reuse distances by analyzing the program structure and memory access patterns at compile time. Early work by Beyls and D’Hollander~\cite{Beyls_locality} applied reuse distance analysis on loop nests using abstract interpretation. Xue~\cite{xue1997reuse} and Kandemir~\cite{kandemir_compilerSupport} proposed compiler techniques that approximate data locality using symbolic loop analysis. More recently, Ming Ling et al.~\cite{Ling_fast_modeling_l2} introduced a hybrid approach that uses software traces and reuse-sensitive trees (RST) to model the behavior of the L2 cache reuse distance. However, this technique still partially depends on the run-time information. Another notable approach, CUDAsap~\cite{CUDAsap}, calculates basic block execution counts at compilation time.

Most static methods either rely on simplifying assumptions or are limited to small, statically analyzable code segments. The existing approaches are so far unable to robustly handle nested loops by changing the array references and predicting reuse histograms from there. Moreover, many hybrid or semi-static techniques still depend on profiling certain parameters at runtime, which limits their usability in fully static compilation pipelines.

\vspace{0.5em}

In this work, we address these limitations by proposing a fully static method tailored for array references in nested loops, capable of estimating reuse-distance histograms with no dependency on execution or trace collection. By leveraging smaller loop bounds, structural memory access patterns, index expressions, and depending on mathematical relations, our approach provides a generalizable alternative to dynamic reuse profiling, enabling faster and scalable prediction of memory behavior suitable for integration into compilers or early design tools. This work is closely aligned with our previous study~\cite{memsys24_paper}. However, it presents significant enhancements over the earlier approach. The key improvements include:
\begin{itemize}
\item \textbf{Loop block specific calculation}: Introduces a separate calculation process for each loop block, which was absent in the earlier work. In this approach, loop blocks are isolated, the prediction process is applied individually to each, and the results are merged at the end with other loop and non-loop blocks. This feature enables more accurate predictions for larger and more complex programs.
\item \textbf{Loop annotation and array index tracking}: Integrates techniques to identify and annotate loop-controlling variables for multidimensional array accesses, and to compute shared array references used across multiple loop blocks, including the handling of cold misses.
\item \textbf{Validation across complex array based applications}: Achieves high prediction accuracy for both RDH and cache hit rates, closely matching PARDA calculations, and demonstrates improved performance on complex problems involving arrays in nested loop environments.
\end{itemize}



\section{Methodology}
\label{sec:method}
\begin{figure}[ht]
    \centering
    \includegraphics[trim=0mm -5mm 0mm 0mm, clip, width=.35\textwidth]{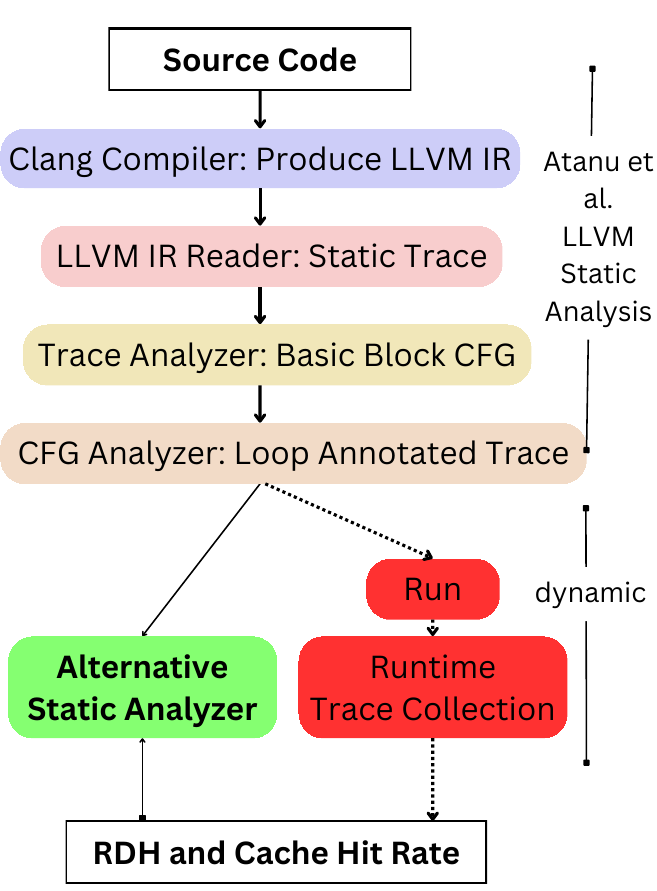}
    \vspace{-5mm}
    \caption{Steps of static analysis-based reuse profile and cache hit rate prediction from the source code.}
    \label{fig:overview}
\end{figure}

We aim to statically predict the RDH of array references in programs with nested loops, without executing the program or collecting memory traces. Although the high-level structure is shown in Figure~\ref{fig:overview}, this work focuses mainly on developing the Alternative Static Analyzer. This model takes input of the \texttt{loop annotated trace}, which is an output of the Atanu et al.~\cite{LLVM_Atanu} developed LLVM static analyzer. After the necessary calculations, it produces the histogram of the reuse profile and the cache hit rates. For better understanding, we use the example program in Figure~\ref{fig:example_program} as a running example to describe each step.

\begin{figure}[t]
    \centering
    \lstset{style=codeListStyle}
    \begin{lstlisting}[language=C,numbers=left]    
for (i = 0; i < 100; i++)
  for (j = 0; j < 200; j++) {
    tmp[i][j] = 0;
    for (k = 0; k < 300; ++k)
        tmp[i][j] += alpha * A[i][k] * B[k][j];
  }

for (i = 0; i < 150; i++)
  for (j = 0; j < 250; j++) {
    D[i][j] *= beta;
    for (k = 0; k < 350; ++k)
      D[i][j] += tmp[i][k] * C[k][j];
  }
    \end{lstlisting}
    \caption{Example code of nested loops and arrays.}
    \label{fig:example_program}
\end{figure}

\begin{figure}[htb]
    \centering
    \includegraphics[trim=0mm 0mm 0mm 0mm, clip, width=.48\textwidth]{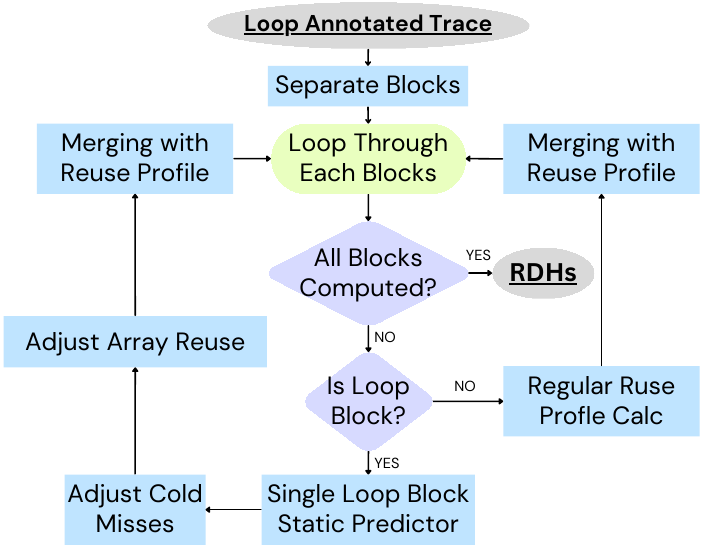}
    \caption{Workflow of the Alternative Static Analyzer.}
    \label{fig:workflow}
\end{figure}

After receiving the Loop Annotated Trace at ~\ref{subsec:loop_annotated_trace}, we separate the trace into various blocks. Figure~\ref{fig:workflow} shows the workflow of the Alternative Static Analyzer. Each loop block can be marked with the starting square braces and numbers. The loop blocks are calculated differently from a regular block. Each loop block is processed with the Static Predictor in isolation as described in ~\ref{subsec:single_loop_block}. The single-loop block calculation reduces the problem size by setting smaller loop bounds and then predicting the actual loop bounds at the end. Then we adjust or exclude the redundantly calculated cold misses in ~\ref{subsec:adjusting_cold_misses}. Some arrays might be used in multiple loop blocks where their reuses need to also be reported, which is covered in ~\ref{subsec:adjusting_array_reuses}. After adjusting those reuse scenarios, the rescue profiles are merged as explained in ~\ref{subsec:merging_reuse_profile}. When all the blocks are calculated, the final RDH appears as a result of our Static Analyzer.

\begin{figure}[t!]
    \centering
    \lstset{style=codeListStyle}
    \begin{lstlisting}
['retval', 'alpha', 'beta', 'i', '[100', 'i', 'j', '[200', 'j', 'i', 'j', 'tmp_array-i-j', 'k', '[300', 'k', 'alpha', 'i', 'k', 'A_array-i-k', 'k', 'j', 'B_array-k-j', 'i', 'j', 'tmp_array-i-j', 'tmp_array-i-j', 'k', 'k', ']', 'k', 'j', 'j', ']', 'j', 'i', 'i', ']', 'i', 'i', '[150', 'i', 'j', '[250', 'j', 'beta', 'i', 'j', 'D_array-i-j', 'D_array-i-j', 'k', '[350', 'k', 'i', 'k', 'tmp_array-i-k', 'k', 'j', 'C_array-k-j', 'i', 'j', 'D_array-i-j', 'D_array-i-j', 'k', 'k', ']', 'k', 'j', 'j', ']', 'j', 'i', 'i', ']', 'i']

    \end{lstlisting}
    \caption{Generated loop annotated trace from the source code.}
    \label{fig:loop_annotated_trace}
\end{figure}

\subsection{Loop Annotated Trace (Input)}
\label{subsec:loop_annotated_trace}
We generate the loop annotated trace using the LLVM static analyzer~\cite{LLVM_Atanu}. Their method is completely static and is based on compile-time information. To generate a loop-annotated trace from source code, the C program is compiled to LLVM Intermediate Representation (IR) using Clang with -emit-llvm, producing a human-readable IR file. This IR is then parsed to extract basic block and control flow information, which is used to construct a Control Flow Graph (CFG) and estimate basic block execution counts based on branch probabilities and input sizes. Finally, they analyze the most probable execution paths within the CFG to identify loop boundaries and use GEP instructions to annotate array index variables and loop controllers, resulting in a static trace enriched with loop and memory access context. After following their process, the output trace looks like Figure~\ref{fig:loop_annotated_trace}. This trace is the input to our Static Analyzer.

\subsection{Separating Blocks}
From the Loop Annotated Trace, we generate separate blocks in this step. This helps the analyzer to understand which blocks are related to an array execution. This is because there is a separate calculation process for loop blocks. After executing the steps, the blocks look the same as depicted in Figure~\ref{fig:separate_blocks}. It can be seen that out of 5 blocks, the second and fourth blocks are the loop blocks and need to be calculated differently as a Single Loop Block Static Predictor~\ref{subsec:single_loop_block}.
\begin{figure}[tbh]
    \centering
    \lstset{style=codeListStyle}
    \begin{lstlisting}
['retval', 'alpha', 'beta', 'i']
['[100', 'i', 'j', '[200', 'j', 'i', 'j', 'tmp_array-i-j', 'k', '[300', 'k', 'alpha', 'i', 'k', 'A_array-i-k', 'k', 'j', 'B_array-j-k', 'i', 'j', 'tmp_array-i-j', 'tmp_array-i-j', 'k', 'k', ']', 'k', 'j', 'j', ']', 'j', 'i', 'i', ']']
['i', 'i']
['[150', 'i', 'j', '[250', 'j', 'beta', 'i', 'j', 'D_array-i-j', 'D_array-i-j', 'k', '[350', 'k', 'i', 'k', 'tmp_array-i-k', 'k', 'j', 'C_array-j-k', 'i', 'j', 'D_array-i-j', 'D_array-i-j', 'k', 'k', ']', 'k', 'j', 'j', ']', 'j', 'i', 'i', ']']
['i']
    \end{lstlisting}
    \caption{Separated Blocks from the Loop Annotated Trace.}
    \label{fig:separate_blocks}
\end{figure}

\subsection{Single Loop Block Static Predictor}
\label{subsec:single_loop_block}
\begin{figure}[tbh]
    \centering
    \includegraphics[width=.49\textwidth]{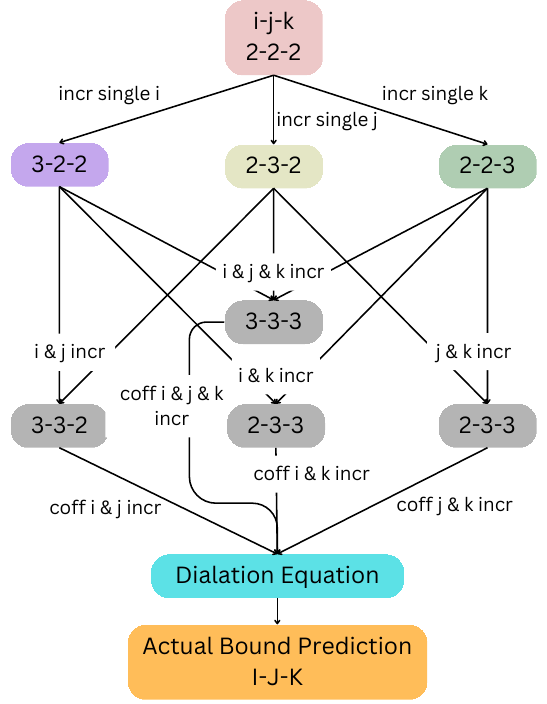}
    \caption{Estimating Reuse Profile from Smaller Problem Sizes. We analyze small problem sizes to derive predictive coefficients. We systematically vary each loop bound (i, j, k) individually and in combination to isolate its effect on the reuse profile, which informs the terms of the Dilation Equation.}
    \label{fig:loop_block_dialation_prediction}
\end{figure}

To predict the reuse profile for large loop bounds of a single loop block (e.g., 100-200-300 bounds in the first loop block), we begin by analyzing with smaller bounds. Figure~\ref{fig:loop_block_dialation_prediction} depicts this process. We replaced the bigger loop bounds with smaller numbers such as 2-2-2, 2-2-3, 2-3-2, 3-2-2, 2-3-3, 3-2-3, 3-3-2, and 3-3-3. For each smaller loop-bound problem, we unfold the loop bounds and calculate the reuse profile for every smaller loop-bound example. In the first step, we observe the impact of incrementing one loop variable at a time to observe its effect on the reuse profile. For example, comparing 2-2-2 to 2-2-3 reveals the influence of increasing \texttt{k} by one, while moving from 2-2-2 to 2-3-2 highlights the impact of increasing \texttt{j} by one. Next, we examine coefficients where two variables are increased simultaneously, such as 2-3-3, to capture the combined effects of both \texttt{j} and \texttt{k} incrementing by one. In this way, we mark the coefficients or multipliers for each incremental situation. This step-by-step process allows us to observe and quantify how reuse patterns evolve with changes in loop bounds. Using the differences observed across these smaller cases, we derive coefficients that represent the contribution of each loop dimension to the reuse behavior. These coefficients are then incorporated into the dilation equations.

For the dilation equation, as this particular problem involves loops with three nesting levels, we categorize the problem as a third-degree, three-variable problem since the loop depth is three and the loop variables are
$i$, $j$, $k$. Equation~\ref{eq:equation_3_degree} illustrates the equation used to model the example scenario.

\begin{align}
\label{eq:equation_3_degree}
    Frq &= B_{222} + Dist_I \times Incr_I + Dist_J \times Incr_J + Dist_K \times Incr_K \notag \\
    &\quad + Coff_{IJ} \times Incr_I \times Incr_J
    + Coff_{JK} \times Incr_J \times Incr_K \notag \\
    &\quad + Coff_{IK} \times Incr_I \times Incr_K  + Coff_{IJK} \times Incr_I \times Incr_J \times Incr_K
\end{align}

However, if the problem employs two nested loops, we apply Equation~\ref{eq:balance} to compute the reuse distance frequency.

\begin{equation}
\label{eq:balance}
    Frq= B_{22}  + Dist_J \times Incr_J + Dist_K \times Incr_K + Coff_{JK} \times Dist_J \times Dist_K 
\end{equation}

In this way, we predict the frequency for the larger bound by applying the required dilation equation based on the loop depth and the variables involved. This technique represents our model as a problem-size independent solution while completely being static without executing the full-scale program.

\subsection{Adjusting Cold Misses}
\label{subsec:adjusting_cold_misses}
Since our approach analyzes and predicts each loop block independently, the same array can appear in multiple loop blocks. As a result, the reuse profile for each block might classify the same array references as cold misses, indicating that the reference is being seen for the first time. However, a cold miss should only be recorded when a memory location is accessed for the first time across the entire execution, not per loop block. Therefore, when combining reuse profiles from different loop blocks, we must identify and eliminate those redundant cold misses to avoid overestimating the total number of unique accesses. This adjustment ensures that cold misses are reported only once, when the reference truly appears for the first time, maintaining the accuracy of the overall RDH. For example, Figure~\ref{fig:redundant_cold_misses} explains such an example in which the array tmp is introduced to the memory stack by the first loop block for references $0-0$ to $99-299$. When the second loop block is calculated independently, it marks $tmp[0][0]$ to $tmp[149][349]$ as cold misses. However, $tmp[0][0]$ to $tmp[99][299]$ have been introduced before by the first loop block, and hence those 30,000 redundant cold miss frequencies are deducted.

\begin{figure}[tbh]
\centering
\begin{minipage}[t]{0.5\textwidth}
\centering
\begin{tabular}{cccccc}
0-0 & 0-1 & 0-2 & 0-3 & ... & 0-299\\
1-0 & 1-1 & 1-2 & 1-3 & ... & 1-299\\
2-0 & 2-1 & 2-2 & 2-3 & ... & 2-299\\
3-0 & 3-1 & 3-2 & 3-3 & ... & 3-299\\
... & ... & ... & ... & ... & ...\\
99-0 & 99-1 & 99-2 & 99-3 & ... & 99-299\\
\end{tabular}
\caption*{a. tmp[0][0] to tmp[99][299] introduced by the first loop block}
\end{minipage}

\hfill

\begin{minipage}[t]{0.5\textwidth}
\centering
\begin{tabular}{cccccccc}
0-0 & 0-1 & 0-2 & 0-3 & ... & 0-299 & ... & 0-349\\
1-0 & 1-1 & 1-2 & 1-3 & ... & 1-299 & ... & 1-349\\
2-0 & 2-1 & 2-2 & 2-3 & ... & 2-299 & ... & 2-349\\
3-0 & 3-1 & 3-2 & 3-3 & ... & 3-299 & ... & 3-349\\
... & ... & ... & ... & ... & ... & ... & ...\\
99-0 & 99-1 & 99-2 & 99-3 & ... & 99-299 & ... & 99-349\\
... & ... & ... & ... & ... & ... & ... & ...\\
149-0 & 149-1 & 149-2 & 149-3 & ... & 149-299 & ... & 149-349\\
\end{tabular}
\caption*{b. tmp[0][0] - tmp[99][299] again introduced by the second loop}
\end{minipage}

\caption{Redundant cold misses from shared array usage across loop blocks}
\label{fig:redundant_cold_misses}
\end{figure}

\subsection{Adjusting Array Reuses}
\label{subsec:adjusting_array_reuses}
When an array is first accessed in an earlier loop block, all its references are introduced through the loop variable values of that first loop block. If the same array references appeared in later loop blocks, those references should find a reuse distance in the following loop blocks, even though the access patterns might be infrequent. To avoid overestimating reuse distances, we apply sequence-based predictions that consider the order and timing of accesses across loop blocks. This adjustment reports reuses of the same array references that are used in multiple loops, leading to more accurate profiling. For example, in Figure~\ref{fig:redundant_cold_misses}, all memory references from $tmp[0][0]$ to $tmp[99][299]$ find reuses in the second loop block.

\subsection{Merging Reuse Profile}
\label{subsec:merging_reuse_profile}
After statically analyzing each loop block individually, we obtain a RDH and a list of cold misses corresponding to first-time memory accesses within each block. However, to construct a final program-wide reuse profile, it is essential to merge these partial results into a unified histogram while accounting for overlapping memory references. Our merging strategy involves aggregating the RDHs by summing the frequency counts of the matching reuse distance keys across all blocks. More importantly, we resolve redundant cold misses by tracking memory references seen in earlier subsections. If a memory reference is marked as a cold miss in a later block, even though it has already appeared in any previous block, it is excluded from the global cold miss list. This ensures that the final RDH and cold-misses profile accurately reflect the temporal locality across the full execution context of the program, rather than overstating cold starts due to per-block isolation. After the merging process is complete, the cold miss list contains each unique memory reference only once.

\begin{figure}[tbh]
    \centering
    \includegraphics[trim=0mm 0mm 0mm 0mm, clip, width=.49\textwidth]{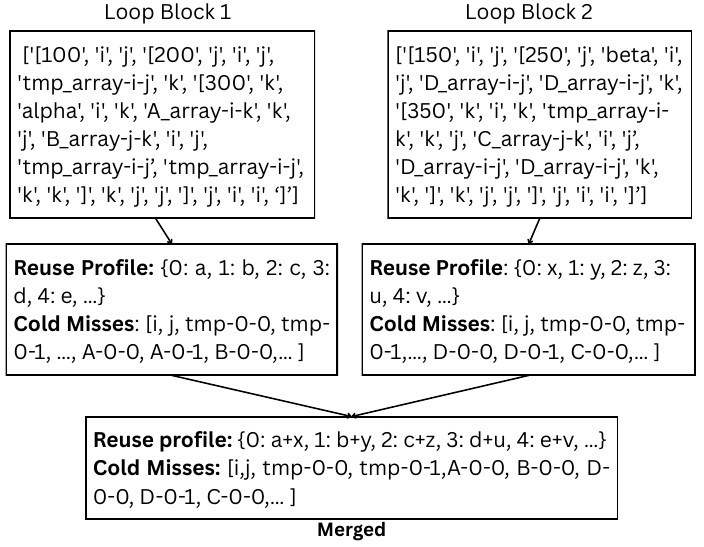}
    \caption{Merging reuse profile and cold misses.}
    \label{fig:merging_reuse_profile}
\end{figure}

Figure~\ref{fig:merging_reuse_profile} demonstrates the merging technique. After each block calculates its reuse profile, they are merged at the end. The frequencies of the same reuse distance are added, for instance, the reuse distance 0, both frequency a in block 1 and frequency x in block 2 are added up at merging. In the same way, while merging the cold misses, only the unique references are kept in that list.
After merging all the blocks, the final reuse profile is reported as the output, and all cold misses are reported as the frequency to the -1 key.

\section{Results}
\label{sec:results}
This section summarizes our experimental results and compares them with the reference dynamic tools. Our static approach predicts reuse profiles and cache hit rates very close to the reference model, showing a significant speedup in calculation time. 

Overall, our results highlight a fundamental trade-off; our static approach perfectly captures the high-frequency, short-distance reuses that dominate the behavior of smaller, tighter caches, while a dynamic approach retains perfect fidelity for long-distance reuses that only matter in very large caches. 

\subsection{Comparison Methods \&  Metrics}
\subsubsection{Reference Model: PARDA}
To evaluate the precision and performance of our static reuse profiling approach, we use {PARDA}~\cite{PARDA:Niu} as a reference dynamic model. PARDA utilizes a tree-based structure with a complexity of $O(n \log n)$ and performs a parallel computation to calculate reuse distances from collected memory traces, providing both high accuracy and scalability. We compare our statically predicted RDH and cache hit rates against PARDA's dynamically calculated output. The execution times are reported from the time of starting the compilation of the source code to the completion of the RDH calculation in both models. The comparison demonstrates the efficiency of our static approach over trace-driven methods.

\subsubsection{{Cache}}
\label{subsubsec:cache-ref}
We evaluated our static prediction method in various cache sizes of 32KB, 256KB, and 1MB, using a fixed associativity of 8 and a line size of 64 bytes. The results show that our model accurately captures reuse behavior across varying cache capacities, closely matches the reference dynamic profiling tool, and demonstrates scalability for different memory hierarchies.

\subsubsection{{Applications Used for Validation}}
We have used multiple applications from PolyBench~\cite{polybenchc} benchmark suite. Table~\ref{tab:kernels} lists the benchmark applications we tested. Each application may contain two or three nested loops. The 2mm application is used for problem scaling analysis across four different datasets, whereas the other applications are used to compare their behavior in our model while maintaining a constant problem size in the small configuration.

\subsubsection{{Cache Hit Rate Measurement}}
We use an analytical memory model known as the Stack Distance-based Cache Model (SDCM)~\cite{brehob:analytical} to estimate cache hit rates. This model accepts the reuse profile of a program along with cache parameters and computes the corresponding hit rate. For a fair comparison, we keep the cache configuration constant for both our model and the reference model. SDCM uses the following equation~\ref{eq:phit} to calculate the probability of a hit {\em P(h)} for the entire reuse profile.

\begin{equation}
\label{eq:phit}
P(h) = \sum\limits_{i=0}^N P({D_i}) \times P({h\mid D_i})
\end{equation}

\noindent where, $P({D_i})$ is the probability of $i^{th}$ reuse distance ($D$) in a reuse distribution and ($P(h\mid D)$) is the conditional probability of hitting given a distance $D$. $P({D_i})$ is calculated from the frequency of $D$ in the reuse histogram. 

\subsection{Problem Scaling Analysis}
In this section, we analyze the effect of increasing problem size, more specifically, loop bounds, while keeping the application same, to observe the performance characteristics. This comparison helps to evaluate how this tool performs as the dataset scales from smaller to larger sizes compared to the reference model. Provides valuable insights into the tool’s scalability and efficiency across varying problem sizes.

\subsubsection{{Dataset}}
We have used the 2mm application and progressively scaled the dataset from Mini to Large. Table~\ref{tab:2mm_loop_array_table} outlines the detailed sizes of the problem. NI, NJ, NK and NL are the loop bounds of the program and they also dictate the array sizes. As the loop bounds increase across different runs, both the execution time and the number of memory references increase significantly. This scaling allows us to evaluate how well our model adapts to larger computational loads.



\begin{table*}[htb]
\centering
\begin{tabularx}{\textwidth}{|c|c|X|}
\hline
\textbf{Dataset} & \textbf{Loop Bounds} & \textbf{Array Dimensions} \\ \hline
Mini   & NI=16, NJ=18, NK=22, NL=24  & tmp[16][18], A[16][22], B[22][18], C[18][24], D[16][24] \\ \hline
Small  & NI=40, NJ=50, NK=70, NL=80  & tmp[40][50], A[40][70], B[70][50], C[50][80], D[40][80] \\ \hline
Medium & NI=180, NJ=190, NK=210, NL=220 & tmp[180][190], A[180][210], B[210][190], C[190][220], D[180][220] \\ \hline
Large  & NI=300, NJ=320, NK=400, NL=420 & tmp[300][320], A[300][400], B[400][320], C[320][420], D[300][420] \\ \hline
\end{tabularx}
\caption{Problem size scaling configurations for the 2mm application.}
\label{tab:2mm_loop_array_table}
\end{table*}

\subsubsection{{Reuse Distance Comparison}}
\begin{figure*}[htbp]
    \centering
    \begin{minipage}{0.49\textwidth}
        \centering
        \includegraphics[width=\textwidth]{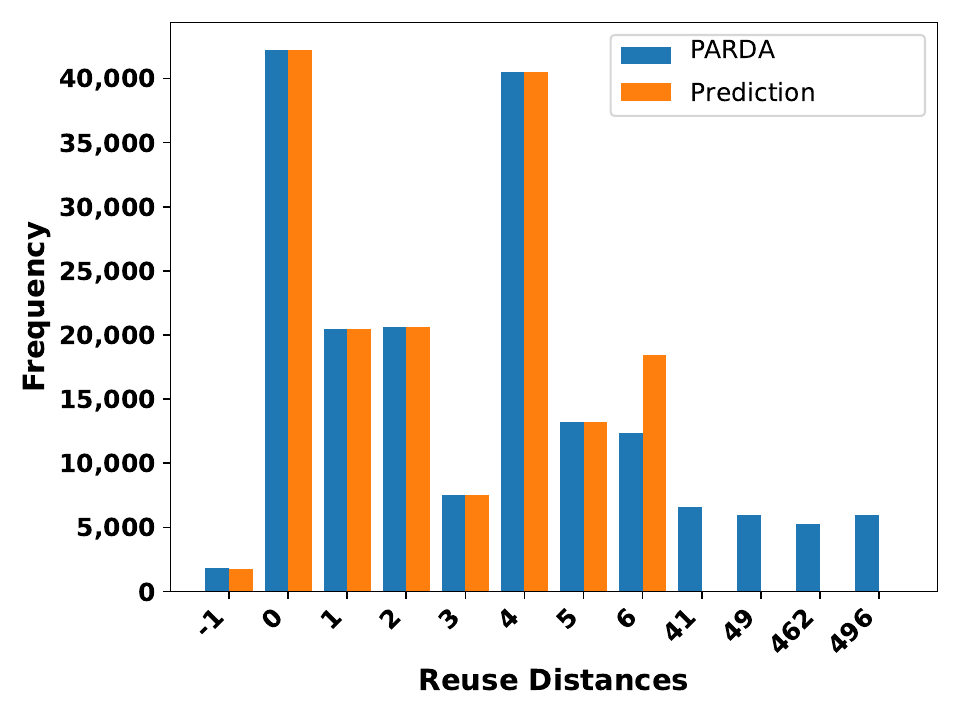}
        \vspace{-6mm} 
        \captionof{subfigure}{Mini}
        \label{fig:mini}
        \vspace{6mm} 
        
    \end{minipage}
    \hfill
    \begin{minipage}{0.49\textwidth}
        \centering
        \includegraphics[width=\textwidth]{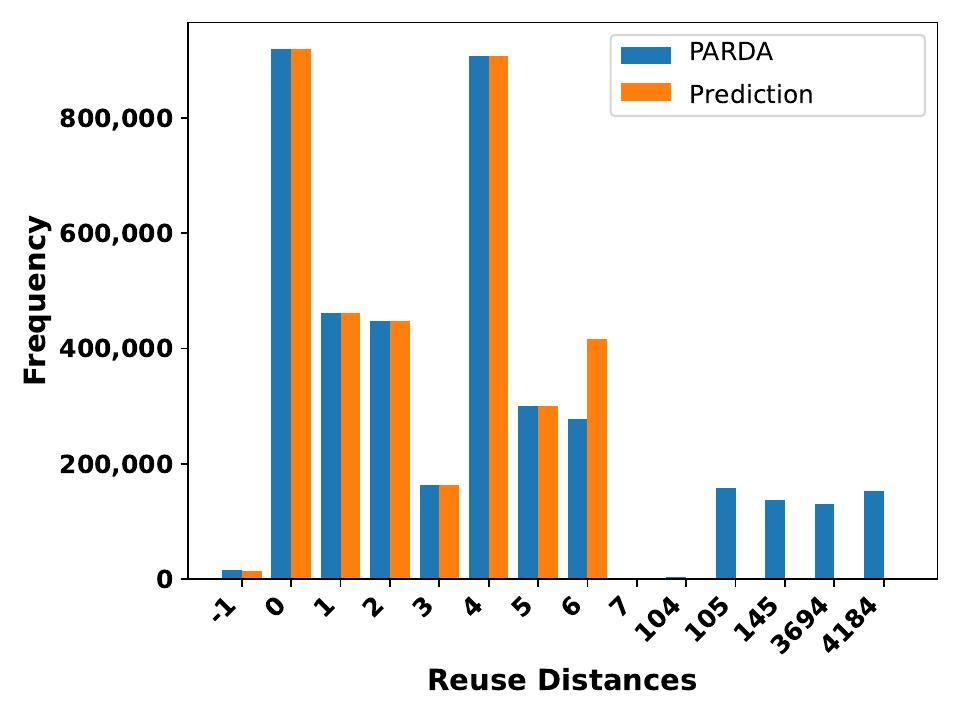}
        \vspace{-6mm} 
        \captionof{subfigure}{Small}
        \label{fig:small}
        \vspace{6mm} 
    \end{minipage}

    \begin{minipage}{0.49\textwidth}
        \centering
        \includegraphics[width=\textwidth]{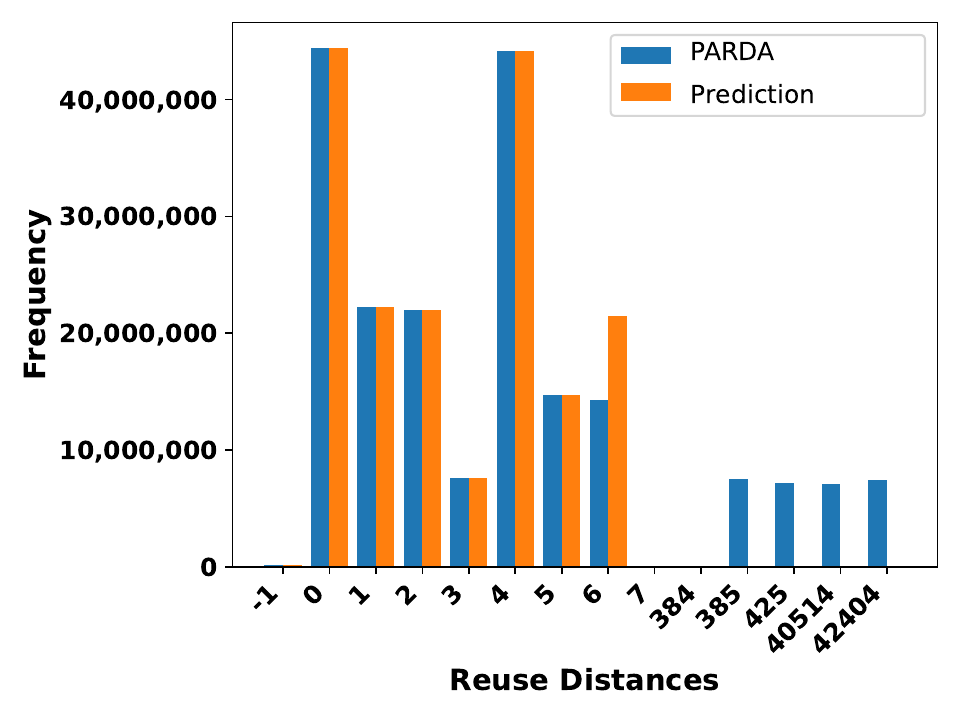}
        \vspace{-6mm} 
        \captionof{subfigure}{Medium}
        \label{fig:medium}
        \vspace{6mm} 
    \end{minipage}
    \hfill
    \begin{minipage}{0.49\textwidth}
        \centering
        \includegraphics[width=\textwidth]{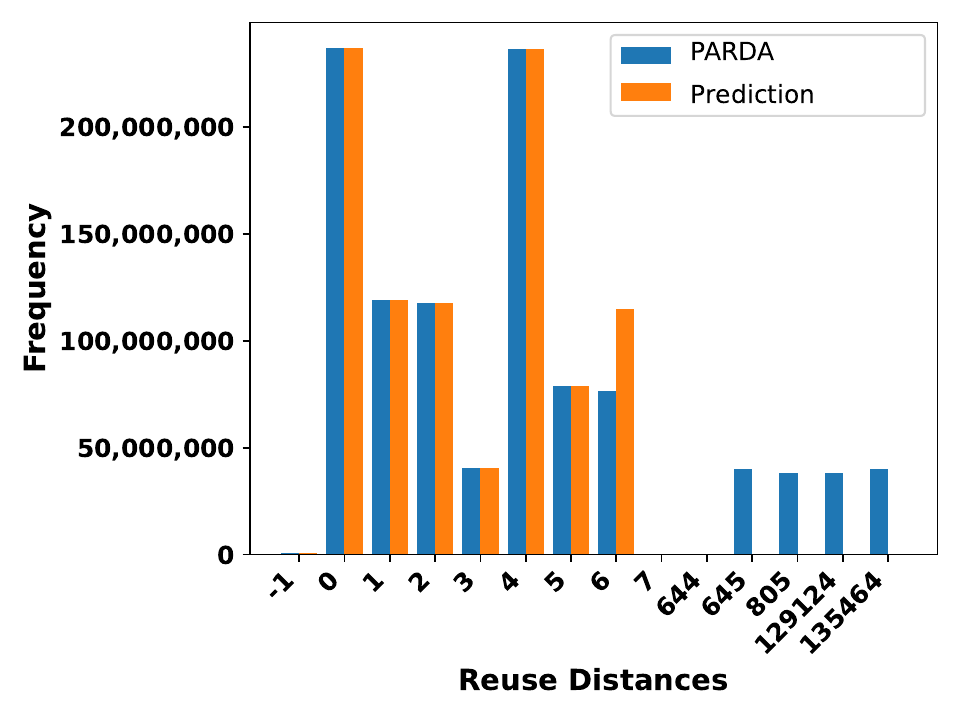}
        \vspace{-6mm} 
        \captionof{subfigure}{Large}
        \label{fig:large}
        \vspace{6mm} 
    \end{minipage}
    \vspace{-6mm} 
    \caption{Reuse Distance Histogram comparison on problem size scaling between the dynamic tool \& our static method.}
    \label{fig:reuse_comparison_scaling}
\end{figure*}

As we compute the reuse profiles, it is essential to note that the application logic remains unchanged; only the loop bounds are scaled accordingly. Figure~\ref{fig:reuse_comparison_scaling} shows the comparison of the frequency of reuse distances across different problem sizes, where the blue bars represent the calculated frequencies of the reference model, and the orange bars indicate the predicted frequency counts of our model. Since a large number of reuse distances are reported, it is difficult to display all of them in the graph. Therefore, we only show that the reuse distances with frequency counts greater than 800 (\textit{Freq(RD) > 800}).

Both our model and the reference model exhibit higher frequencies under the same reuse distances as we move from the Mini to the larger problem sizes. Our model performs well in predicting the highest frequencies at lower reuse distances with notable accuracy. However, as a compile-time approach, it lacks access to the exact timing of memory reference occurrences. As a result, it can only estimate frequency distributions and may not precisely predict reuse distances that arise from the dynamic behavior of array references during runtime.

\subsubsection{{Cache Hit Rate}}
\begin{figure}[htb]
    \centering
    \includegraphics[width=0.49\textwidth]{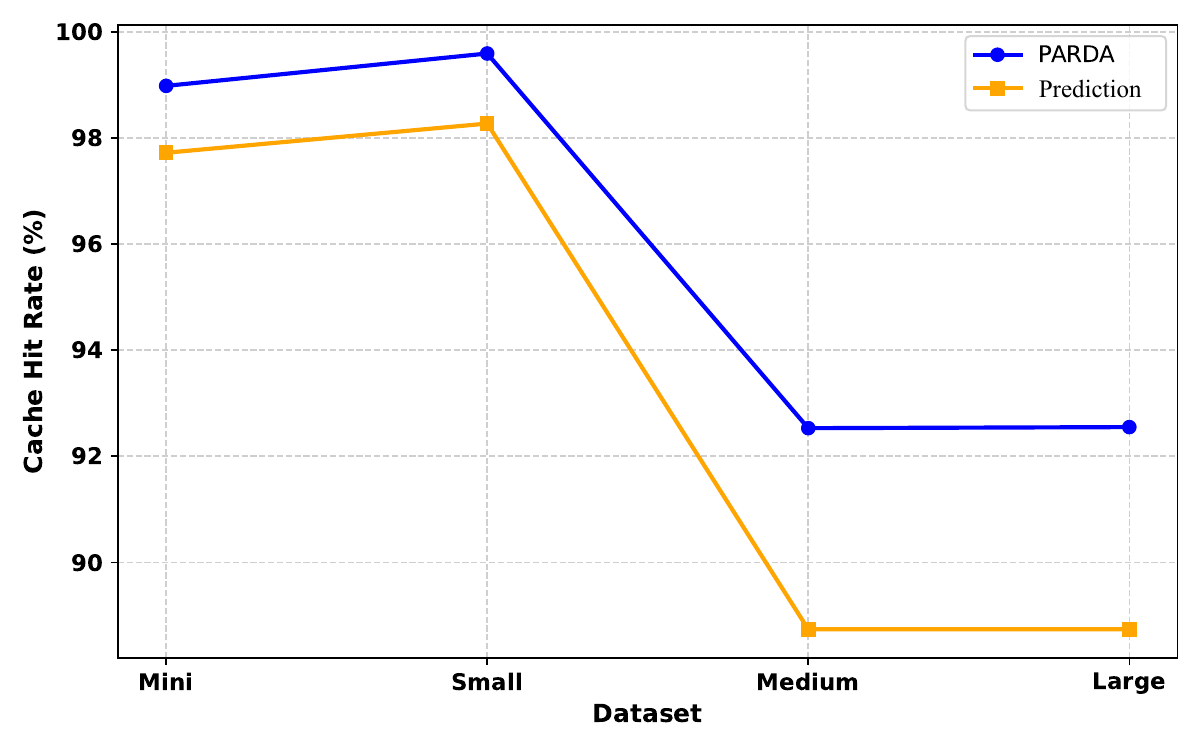}
    \caption{Cache Hit Rate comparison between the reference model (PARDA) and this model across various datasets.}
    \label{fig:cache_hit_scaling}
\end{figure}

We compared the cache hit rates produced by our static model with those from the dynamic PARDA approach across different problem sizes. For this comparison, we keep the cache size 1 Megabyte, block size 64, and 8-way associative. Figure~\ref{fig:cache_hit_scaling} shows the comparison of the cache hit rates between PARDA and our model across various problem sizes. Our model closely approximates the hit rates reported by PARDA for smaller datasets. However, for medium and large problem sizes, the cache hit rate predicted by our model is slightly lower. This discrepancy arises because larger loop bounds introduce more array references, many of which exhibit dynamic behavior that our static analysis cannot fully capture, especially those associated with larger reuse distances, as discussed in Figure~\ref{fig:reuse_comparison_scaling}. In these cases, our model may fail to report reuse distances that fall within the cache, resulting in missed cache hits. In contrast, PARDA calculates dynamically and accurately identifies such reuse distances at runtime and includes them in its hit rate calculation, leading to higher reported accuracy.

\subsubsection{Model Execution Time}
\begin{figure}[htb]
    \centering
    \includegraphics[width=0.49\textwidth]{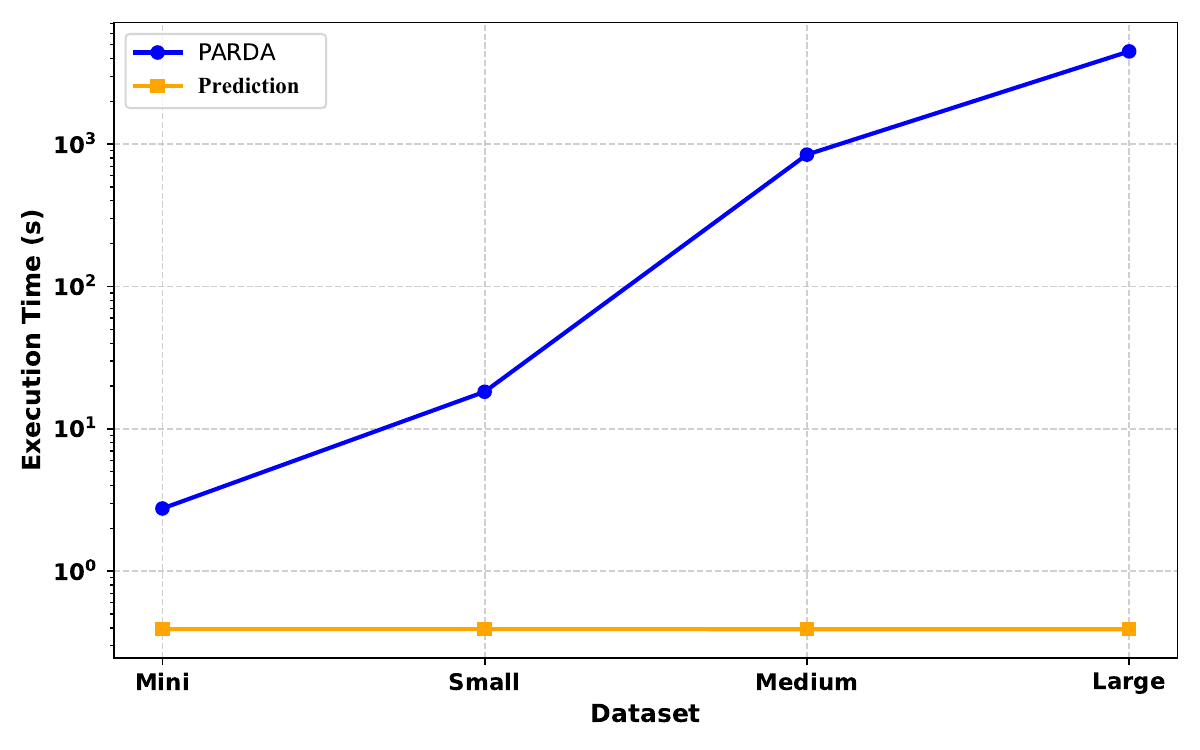}
    \caption{Comparison of Model Execution between PARDA and this model (time from source code to result generation).}
    \label{fig:2mm_execution_time}
\end{figure}

We evaluated the execution time required to generate the RDH from the source program in both the reference model (PARDA) and our proposed model. Figure~\ref{fig:2mm_execution_time} shows the execution time (in seconds) of the different datasets. Our model achieves significant speedups over PARDA, particularly for larger problem sizes.

This significant performance gain is achieved because our model is problem-size independent. Using compile-time information, the model estimates the RDH. Unlike PARDA, it avoids running the program dynamically and eliminates the need for memory trace collection and analysis.

\subsection{Multi-Kernel Evaluation}
In this section, we evaluate the performance of our model using multiple kernels from the PolyBench~\cite{polybenchc} benchmark suite. This analysis helps to understand the effectiveness of this model with other applications.

\subsubsection{{Dataset}}
To further evaluate our model's behavior against the reference model, we tested it on additional PolyBench kernels beyond 2mm. As shown in Table~\ref{tab:kernels}, all the selected applications belong to the linear algebra category. For simplicity, only the kernel portions of these applications were executed. Instead of testing across all data set sizes, we limited our evaluation to the smaller loop-bound configuration described in Table~\ref{tab:2mm_loop_array_table}.

\begin{table}[tbh]
\centering
\begin{tabular}{p{1.6cm} p{0.8cm} c c}
\toprule
\textbf{Applications} & \textbf{Nested Loops} & \textbf{Analysis} & \textbf{Problem Sizes} \\
\midrule
2mm & 3 & Problem Scaling & Mini,\newline Small,\newline Medium,\newline Large \\
3mm & 3 & Multi-Kernel & Small \\
atax & 2 & Multi-Kernel  & Small \\
mvt & 2 & Multi-Kernel & Small \\
gemver & 2 & Multi-Kernel & Small \\
\bottomrule
\end{tabular}
\caption{Other PolyBench kernels used for evaluation.}
\label{tab:kernels}
\end{table}

\subsubsection{{Reuse Distance Comparison}}

\begin{figure*}[htbp]
    \centering
    \begin{minipage}{0.49\textwidth}
        \centering
        \includegraphics[width=\textwidth]{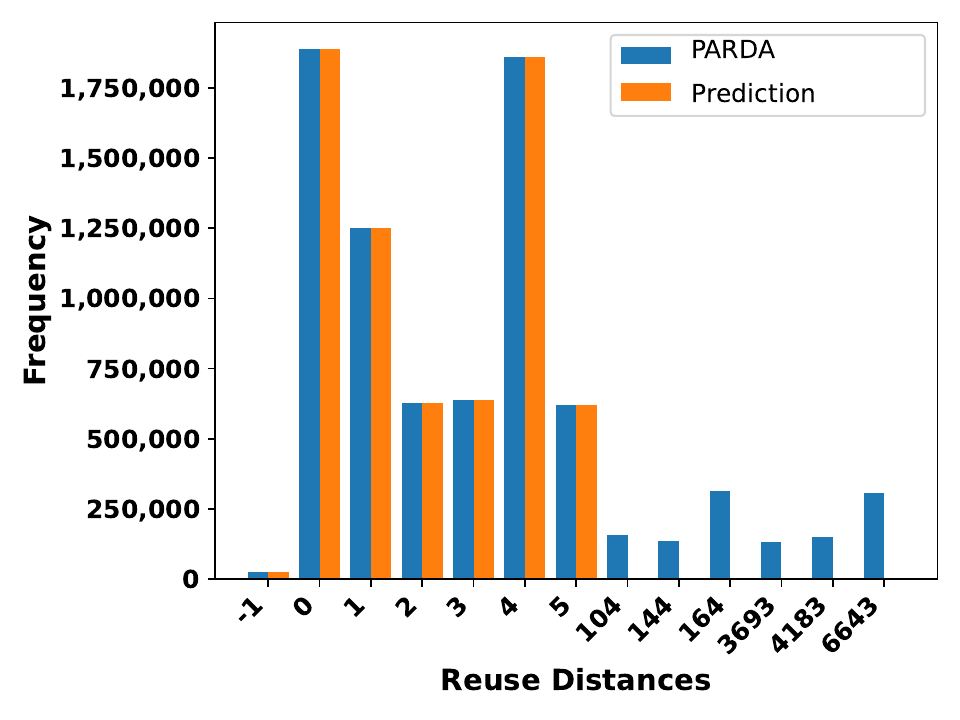}
        \vspace{-6mm} 
        \captionof{subfigure}{3mm}
        \label{fig:rdh_3mm}
        \vspace{6mm} 
        
    \end{minipage}
    \hfill
    \begin{minipage}{0.49\textwidth}
        \centering
        \includegraphics[width=\textwidth]{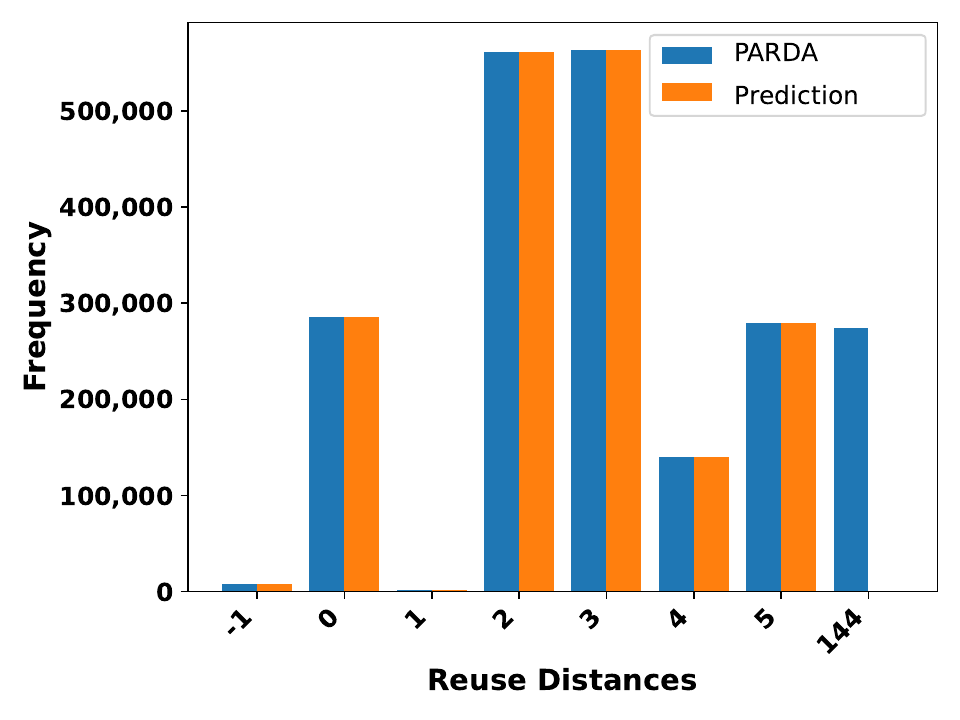}
        \vspace{-6mm} 
        \captionof{subfigure}{atax}
        \label{fig:rdh_atax}
        \vspace{6mm} 
    \end{minipage}

    \begin{minipage}{0.49\textwidth}
        \centering
        \includegraphics[width=\textwidth]{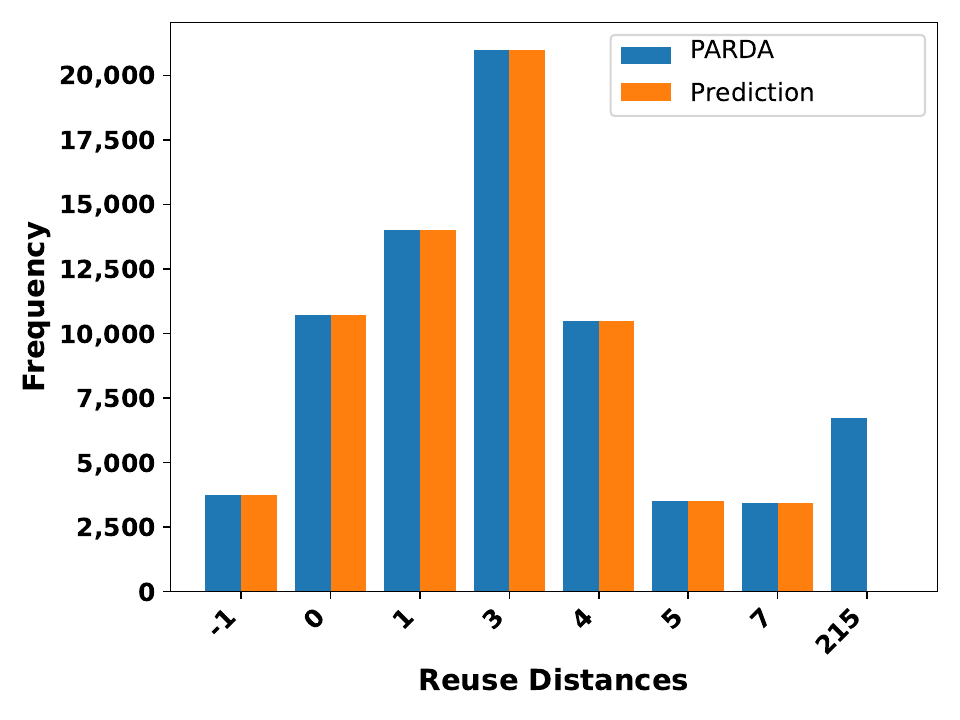}
        \vspace{-6mm} 
        \captionof{subfigure}{bicg}
        \label{fig:rdh_bicg}
        \vspace{6mm} 
    \end{minipage}
    \hfill
    \begin{minipage}{0.49\textwidth}
        \centering
        \includegraphics[width=\textwidth]{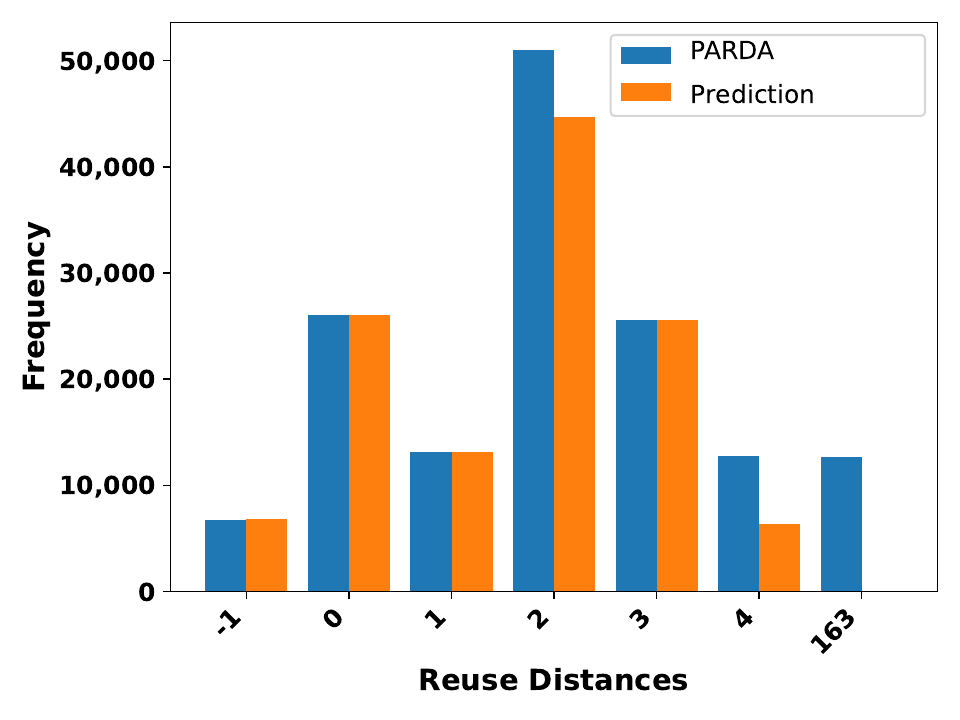}
        \vspace{-6mm} 
        \captionof{subfigure}{mvt}
        \label{fig:rdh_mvt}
        \vspace{6mm} 
    \end{minipage}
    \vspace{-6mm} 
    \caption{Reuse Distance Histogram comparison on different applications.}
    \label{fig:rdh_4_programs}
\end{figure*}

We have run only the kernel applications mentioned in Table~\ref{tab:kernels} and the results are shown in Figure~\ref{fig:rdh_4_programs}.
Once again, our model accurately reports the frequencies for lower reuse distances. However, it struggles to predict reuse distances that are dynamically reported due to the volatility of array accesses. Since we use smaller datasets in this test, kernels such as atax, bicg, and mvt show strong alignment between the reuse distance frequencies of our model and those reported by PARDA. In contrast, 3mm presents a challenge, as it involves multiple arrays with interleaved access patterns that are difficult to estimate without observing the actual runtime behavior.

\subsubsection{{Cache Hit Rate}}
\begin{figure}[htb]
    \centering
    \includegraphics[width=0.49\textwidth]{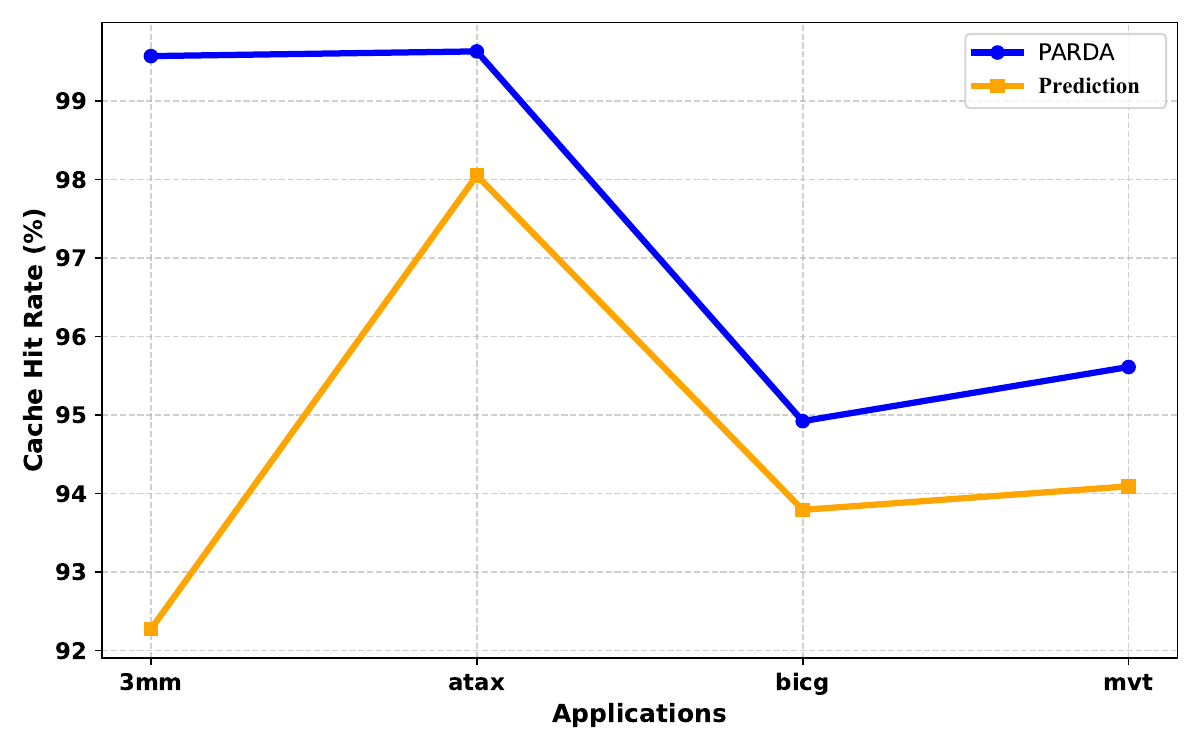}
    \caption{Cache Hit Rate comparison for the applications.}
    \label{fig:cache_hit_apps}
\end{figure}

As shown in Figure~\ref{fig:cache_hit_apps}, our model achieves cache hit rates closely aligned with PARDA in additional benchmarks such as atax, bicg, and mvt. It represents how accurate our predictions are for various applications' memory access patterns. However, 3mm shows a slightly lower hit rate due to its use of multiple arrays with interleaved access, making it difficult for static analysis to capture all reuse opportunities without observing runtime behavior.

\subsubsection{Model Execution Time}
\begin{figure}[htb]
    \centering
    \includegraphics[width=0.49\textwidth]{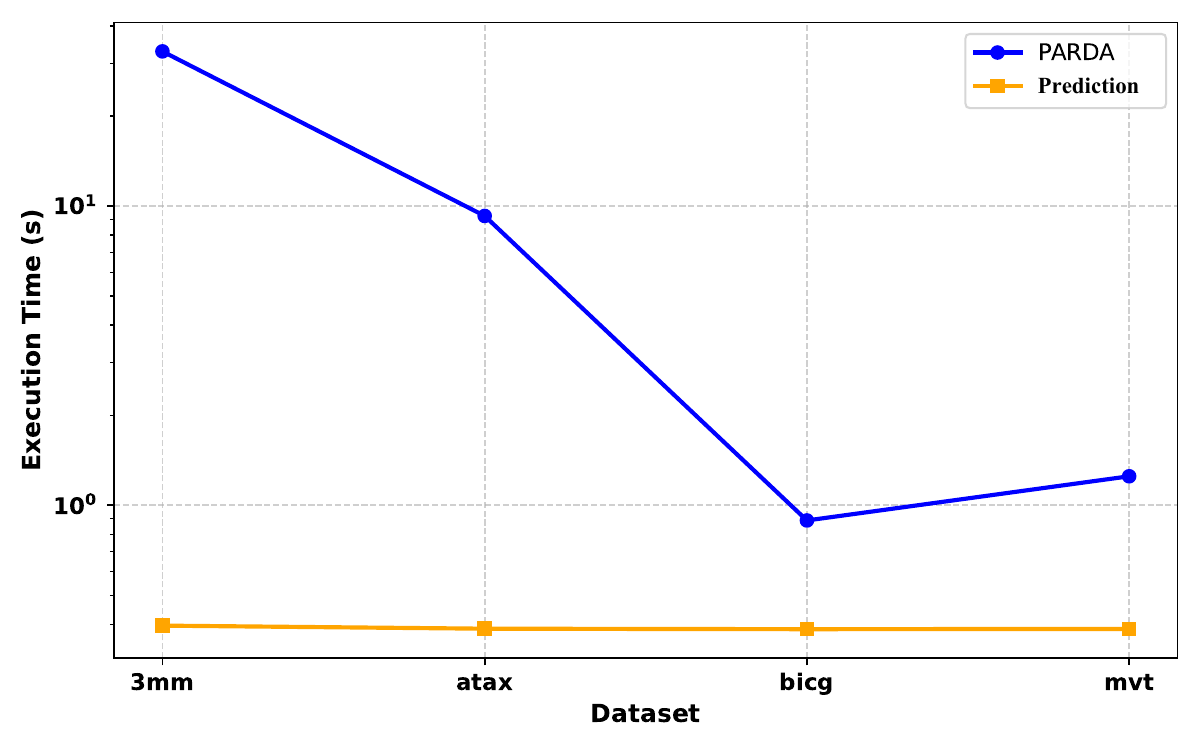}
    \caption{Comparison of Model Execution between PARDA and this model for the benchmark applications.}
    \label{fig:execution_other_apps}
\end{figure}

We compared the execution time of four benchmark applications with the reference model. As previously explained, our model performs static analysis and generates predictions in near-constant time, while PARDA's execution time varies with problem size. For example, as shown in Figure~\ref{fig:execution_other_apps}, the 3mm application involves higher computational complexity and thus takes significantly longer in PARDA. In contrast, bicg is a smaller kernel with a much shorter run-time. Regardless of the application, our model maintains a consistently low execution time. As a result, our static model provides a much faster alternative for estimating the reuse profile without compromising accuracy in critical reuse regions.

\section{Discussion \& Limitations}
\label{sec:limitations}
Although our static model demonstrates strong performance in predicting reuse profiles and cache hit rates efficiently, it does have several limitations:

\begin{itemize}
    \item \textbf{Limited Detection of Long Reuse Distances:} The model currently struggles to identify large and dynamic reuse distances that occur across loop iterations or through indirect memory accesses. These reuse patterns are often captured by a dynamic profiling tool like PARDA but are overlooked in static estimation.
    
    \item \textbf{Accuracy Drops for Large Cache Sizes:} Due to the inability to account for infrequent but reuse distances, the cache hit rate accuracy decreases slightly as the cache size increases.

    \item \textbf{Lacking Branch Predictions:} Our approach does not cover reuse behaviors at function-level reuse behaviors or cases where branches are involved.
    
    \item \textbf{Hand Input Bounds:} Since we cannot execute the program, the loop bounds' values are not available at compile time. Therefore, we manually provide loop exit probabilities before running the static analysis tool.
\end{itemize}

We are currently exploring techniques to address these limitations to better approximate non-local and complex reuse patterns.

\section{Conclusion}
\label{sec:conclusion}
We have presented an almost fully static analysis framework capable of estimating reuse-distance histograms for array-based applications with orders-of-magnitude greater speed than traditional dynamic profilers. We present ``Alternative Static Analyzer'' that estimates RDH and cache hit rates without requiring the execution of the source code or the collection of memory traces. By analyzing the program structure and the reuse patterns at compile time, our model achieves a significant speedup, especially for the larger problems, compared to dynamic tools like PARDA. 
Our approach demonstrates high accuracy in predicting reuse distances and cache hit rates, particularly for frequently accessed references and smaller cache configurations, where it closely matches PARDA’s results. The result indicates that this model is a highly efficient and scalable alternative to the dynamic calculation process for early-stage cache performance analysis. 
However, the current version of the model does not fully capture all array reuse scenarios, especially those involving complex loop interactions or long reuse distances. These limitations lead to an underestimation of cache hit rates at larger cache sizes. We are actively working on addressing these limitations by refining loop analysis and incorporating extended reuse tracking to enhance accuracy across a broader range of memory access patterns.




\section*{Acknowledgment}
The authors thank the anonymous reviewers. Triad National Security, LLC partially funded this research under subcontracts \#581326 and \#C4975. The Department of Energy (DOE) National Nuclear Security Administration (NNSA) under contract DEAC52-06NA25396, and the Los Alamos National Laboratory ASC program under project JAPB-SID1 also partially funded this research. This paper has been approved for public release with LA-UR-25-25754. The opinions, findings, or conclusions expressed in this paper are solely the authors’ and do not necessarily represent those of the DOE or the US government. 

\bibliographystyle{plainurl}
\bibliography{reference}

\end{document}